Robin Quillivic[1,2,3], Salma Mesmoudi[1,2,3,4]

1: EPHE, Ecole Pratiques des Hautes Etudes;  Paris
2: CESSP, CNRS, UMR 8209; Paris
3: ISC PIF, Institut des Systèmes Complexes UAR 3611; Paris
4: Université Paris 1 Panthéon Sorbonne

# La culture du logiciel libre suffit-elle à faire de l'IA un commun?

Une étude de cas sur Hugging Face

# Résumé

Les modèles de langue (LM ou LLM) sont de plus en plus déployés dans le domaine de l'intelligence artificielle (IA) et leurs applications, mais la question qui se pose est de savoir s'ils peuvent être un commun géré et entretenu par une communauté d'utilisateurs. En effet, la domination d'entreprises privées ayant un accès exclusif à des données massives et à des ressources de traitement du langage peut créer des inégalités et des biais dans les LM, ainsi que des obstacles à l'innovation pour ceux qui ne disposent pas des mêmes ressources, nécessaires à leur mise en œuvre. Dans cette contribution, nous examinons le concept de commun et sa pertinence pour penser les LM. Nous soulignons les avantages potentiels de traiter les données et les ressources nécessaires pour créer des LM comme des communs, notamment une accessibilité, une équité et une transparence accrues dans le développement et l'utilisation des technologies d'IA. Enfin, nous présentons une étude de cas centré sur la plateforme Hugging Face, une plateforme open source pour le deep learning conçue pour encourager la collaboration et le partage entre les concepteurs d'IA.

**Mots Clefs:** Intelligence Artificielle, Traitement Automatique du Langage, Economie politique

| Acronyme/ terme | Définition |
|---|---|
| IA / Intelligence Artificielle | Ensemble de méthodes statistiques, de données d'apprentissage et d'infrastructures informatiques qui forment un système capable de réaliser des prédictions ou de répondre à des demandes précises. |
| LM /Modèle de langue | Algorithme statistique qui modélise la distribution de séquences de mots. Un modèle de langue peut par exemple prédire le mot suivant une séquence de mots. Les LMs sont aujourd'hui les briques essentielles de nombreuses applications en traitement automatique du langage. |
| Communs | Ensembles de ressources collectivement gouvernées, au moyen d'une structure de gouvernance assurant une distribution des droits entre les partenaires participant au commun (commoners) et visant à l'exploitation ordonnée de la ressource, permettant sa reproduction à long terme. |

Table 0. Définition des acronymes utilisés

**Introduction**

Depuis plusieurs années, le concept de commun n'est plus seulement l'apanage des universitaires ou des sociologues. En effet, suite au succès de la théorie des communs remis au goût du jour par Elinor Ostrom (Ostrom, 1990); plusieurs mouvements alternatifs et acteurs sociaux s'emparent de cette notion qui est devenue un outil politique permettant de penser un modèle au-delà du marché et de la propriété privée. Les communs numériques et informationnels n'échappent pas à cette tendance et constituent à l'heure actuelle une des formes majeures des nouveaux communs (Alix, 2018). Le logiciel libre en est un bel exemple. En effet, le logiciel libre "représente une forme originale de production et d'organisation du travail" (Alix, 2018) qui repose sur une communauté autonome et autogérée responsable du développement du logiciel et aboutissant à la mise à disposition (gratuite ou payante) d'un logiciel pour lequel tout individu est libre de contribuer et pour lequel le code source est accessible à tous (Stallman, 2002). C'est dans ces communautés que beaucoup d'outils permettant la construction d'algorithmes d'apprentissage et de technologie de l'IA ont été conçus. Ces dernières années, l'IA a connu un essor considérable et est perçu comme un vecteur de changements importants dans la société. En plus d'être parmi les innovations les plus médiatisées du moment, l'IA est considéré par les économies occidentales comme un catalyseur de croissance qui transformera nos économies et nos modes de vie (Agrawal et al., s. d.; Brynjolfsson & McAfee, 2017; Lee et al., 2018). Décrite parfois comme une "General Purpose Technology" (Verdegem, 2022), c'est-à-dire une technologie qui aura un impact sur l'ensemble des économies et des structures sociales. Par exemple, l'IA est souvent citée comme solution pour résoudre des défis tels que le changement climatique (Nordgren, 2023) ou encore les pandémies (Tzachor et al., 2020). Bien que l'IA existe depuis plus de 60 ans et que les périodes d'espoir et d'optimisme ont alterné, "IA winter"[1], il semble que des pièces cruciales du puzzle commencent enfin à se mettre en place. Les avancées spectaculaires concernant les capacités de calculs, la quantité de données disponibles et les progrès concernant l'entraînement des réseaux de neurones ont permis de développer des produits de l'IA (cf figure 0) ayant des cas d'applications et des millions d'utilisateurs quotidiens. Les modèles de langues font partie de cette génération de produits de l'IA et commencent à être accessibles et utilisés par le plus grand nombre (chatGPT, github

---

[1] AI winter - Wikipedia

Copilot[2], Mid-Journey[3]). Pourtant, dans ce nouvel éco-système, nous observons une très forte concentration des pouvoirs de décisions et de gouvernance. Les entreprises ayant les moyens de développer les outils d'IA dernière génération, ne sont que peu nombreuses et contrôlent l'accès aux données, aux ressources de calculs. Il y a une contradiction forte entre la nouvelle ubiquité de ces technologie dans nos vies et le contrôle que nous en avons.

Dans cette contribution, nous nous intéressons particulièrement aux modèles de langues, à leur histoire ainsi qu'à leur évolution future en nous appuyant sur la notion de commun numérique. Nous nous demandons si les communs peuvent permettre de dépasser la contradiction précédente pour des outils tels que les modèles de langue. Quelles seraient les implications pratiques et politiques de ce statut. Dans une première partie, nous rappellerons les principes des communs numériques et montrerons comment les modèles de langues se sont inscrits et ont évolué dans la culture du logiciel libre. Dans un second temps, nous verrons que le changement d'échelle qu'a connu l'IA ces dernières années est une menace pour la gouvernance commune de ces outils. Nous insisterons particulièrement sur la concentration des pouvoirs et moyens de développement au sein des Big-Techs. Enfin dans une dernière partie nous proposons une étude de cas portant sur la jeune pousse américaine Hugging Face en nous demandant si le modèle proposé est en accord avec la définition des communs numériques évoqués plus haut (Table 0).

Figure 0. Schéma de définition d'un produit de l'IA et des ressources associées

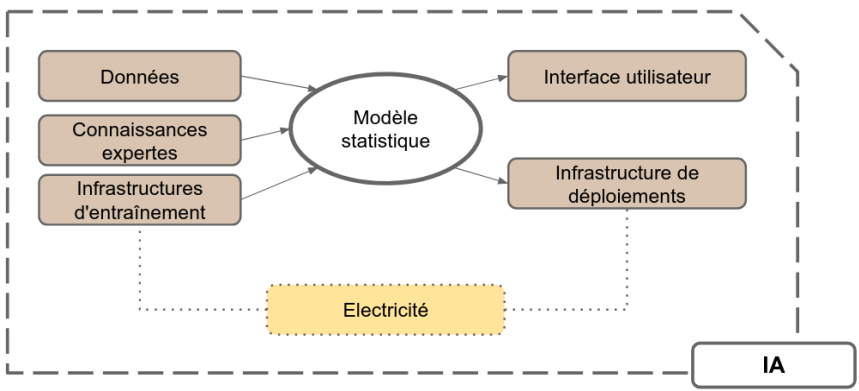

---

[2] [GitHub Copilot · Your AI pair programmer](#)
[3] [Midjourney](#)

1. **L'IA comme "commun", la promesse d'hier**

**Les communs numériques.**
Nous nous plaçons dans le cadre des communs définis par Elinor Ostrom(Ostrom, 1990), les biens communs sont "des ensembles de ressources collectivement gouvernées, au moyen d'une structure de gouvernance assurant une distribution des droits entre les partenaires participant au commun (*commoners*) et visant à l'exploitation ordonnée de la ressource, permettant sa reproduction à long terme." Même si l'ensemble des communs peuvent être considérés comme des communs de la connaissance, car "le partage de la connaissance et de l'information est essentiel à la réussite et à la pérennité de tous les types de communs », les communs numériques se distinguent des "communs naturels", car leur objectif n'est pas la préservation de la ressource, mais plutôt la croissance et la diffusion de celle-ci. Un exemple remarquable de commun numérique fonctionnel est l'encyclopédie Wikipédia. Lancée en janvier 2001, elle propose à l'heure actuelle (2023) des articles rédigés en 315 langues, avec une participation quotidienne de plus de 100 000 personnes[4]. Ces articles sont librement accessibles (sous réserve d'une connexion Internet) et sont alimentés par une communauté de wikipédiens suivant un ensemble de règles et formant leur propre système de gouvernance (Vandendorpe, 2008). D'un point de vue juridique, Wikipédia est soumis à la licence CC-BY-SA[5] qui consacre et protège son statut de bien commun, en empêchant par la clause de partage à l'identique que les contenus puissent être privatisés. Ainsi, il est impossible d'acquérir les contenus de Wikipédia dans le but de les exploiter de manière exclusive. Cependant, cela n'empêche pas la proposition de services payants basés sur les ressources de Wikipédia.

Un autre exemple significatif de commun numérique est le système d'exploitation GNU/Linux. Ce projet, lancé dans les années 1980[6] par Richard Stallman et Linus Torvalds, et développé par une communauté mondiale de contributeurs, offre une alternative libre et open-source aux systèmes d'exploitation commerciaux. Basé sur le partage de code source et la collaboration ouverte, GNU/Linux illustre parfaitement la notion de commun numérique en favorisant la croissance continue du système par l'apport collectif de développeurs et d'utilisateurs. L'absence de restrictions draconiennes sur l'usage et la modification du logiciel encourage une diffusion étendue et une adaptation aux besoins spécifiques, tout en préservant son caractère de bien commun.

---

[4] Wikipédia:Statistiques
[5] About CC Licenses - Creative Commons
[6] GNU - Wikipedia

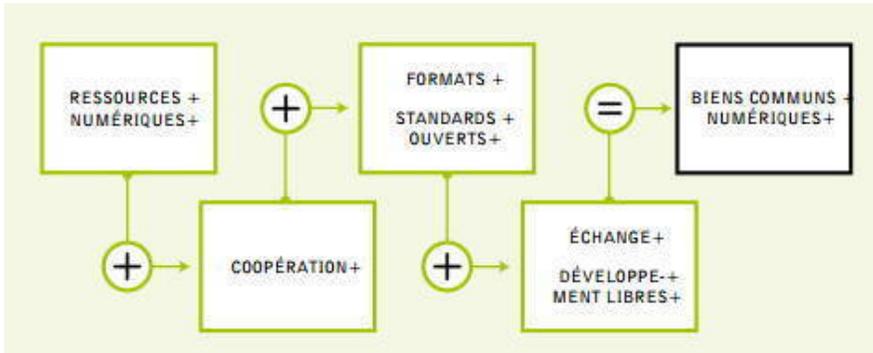

Figure 1. Biens communs numériques selon le rapport – source: biens communs la prospérité par le partage.

Enfin, un commun suppose une ressource partagée et un système de droits et obligations, il se définit aussi par une **gouvernance collective** qui pérennise le commun. Elinor Ostrom et son équipe de chercheurs ont bien remarqué que des droits de propriété, notamment le droit d'exclure, sont nécessaires pour une bonne gouvernance des communs. Wikipédia et Linux s'appuient sur des fondations pour mettre leur gouvernance entre les mains de fondations. Par exemple, c'est la Wikimédia Foundation[7] qui gère le projet avec un budget de plus de 50 millions de dollars. Le conseil d'administration de la Fondation, composé de professionnels avec des compétences pointues en informatique ou en sciences de l'information, définit la stratégie, gère la marque, les projets, etc., comme dans d'autres organisations.

**La culture de l'Open-source dans l'IA.**

L'intelligence artificielle (IA) représente une discipline relativement récente, toutefois elle puise ses racines dans de multiples avancées appartenant à des domaines plus anciens. En effet, pour démocratiser et rendre l'IA accessible, il est impératif en premier lieu de partager et diffuser les connaissances mathématiques qui sous-tendent la compréhension et la conception des programmes d'IA. Parallèlement, il est essentiel de développer des outils et des structures informatiques capables de supporter les calculs complexes nécessaires à l'IA. Enfin, la mise à disposition de grandes quantités de données de qualité est indispensable pour l'entraînement et la reproduction des modèles IA. Ainsi, la complexité inhérente à la mise en œuvre de l'IA a favorisé l'émergence de contributions provenant de diverses communautés. Dans cette partie, nous allons nous arrêter sur trois exemples de projet/communauté qui ont contribué significativement aux développements de l'IA. En commençant par l'échange de connaissances sur Arxiv[8] à l'efficacité de développement avec PyTorch (Paszke et al., 2019), puis en détaillant l'initiative ONNX (Open Neural Network Exchange) qui permet la portabilité des modèles de Deep Learning.

---

[7] Accueil – Fondation Wikimedia

Arxiv, une plateforme en ligne qui permet la diffusion rapide d'articles scientifiques préliminaires non révisés par des pairs, a émergé comme un pilier de la collaboration mondiale en mathématiques et en physique depuis 1991. Les chercheurs y partagent leurs dernières découvertes, facilitant ainsi l'échange d'idées et la progression accélérée des connaissances dans le domaine de l'IA.

PyTorch (Paszke et al., 2019), une bibliothèque open source pour l'apprentissage automatique, s'est imposée comme un outil incontournable. Bien que créé par Meta, PyTorch est depuis 2018 gouverné par la "PyTorch Foundation", une sous-division de la "Linux Foundation"[9].

Le format ONNX a résolu un défi crucial en permettant la portabilité des modèles IA entre différentes plateformes et frameworks de développement. Également sorti, des laboratoires de Facebook, il rejoint la "Linux foundation AI" en 2019[10]. Cette initiative favorise la collaboration en éliminant les obstacles liés à la compatibilité des modèles, accélérant ainsi leur déploiement et leur développement. Enfin, Wikipedia, présenté plus haut, a longtemps servi de source de données ouverte pour le traitement automatique du langage naturel (NLP). Son vaste corpus d'articles fournit un réservoir riche et varié pour l'entraînement des LM, contribuant alors à améliorer la modélisation par des machines du langage humain.

Ces exemples mettent en évidence comment les initiatives open source ont largement contribué à façonner l'IA telle que nous la connaissons aujourd'hui, mais également que certaines entreprises ont fait de l'open source une stratégie de long terme. En effet, l'open source permet d'élargir sa communauté, de faire de sa technologie un standard, de participer à l'éducation des autres développeurs mais également d'attirer des talents.

---

[8] https://arxiv.org/
[9] Linux Foundation - Wikipedia
[10] Open Neural Network Exchange - Wikipedia

**Histoire des modèles de langue**

Les premiers modèles de langues sont apparus il y a plusieurs décennies, avec des jalons importants qui ont façonné leur évolution. Dès les années 1940 et 1950, les premiers systèmes informatiques destinés à la traduction automatique ont été développés dans le contexte de la guerre froide, marquant les débuts de la recherche en traitement automatique du langage naturel (TALN). Cependant, ces premiers efforts étaient souvent le fruit de travaux gouvernementaux et universitaires. Dans les années 1970 et 1980, des systèmes comme SHRDLU (Winograd, 1972) ou ELIZA (Weizenbaum, 1966) ont émergé, permettant aux ordinateurs de comprendre des commandes et des instructions en langage naturel. Les années 1990 ont vu la montée en puissance des moteurs de recherche, avec des entreprises privées telles que AltaVista[11] et Yahoo![12] investissant dans le développement de modèles de recherche sémantique. Cependant, le domaine de l'IA et du TALN a été caractérisé par des progrès lents et fragmentés, avec des initiatives provenant d'un mélange de chercheurs universitaires, d'entreprises privées et d'organisations gouvernementales. Au même moment dans le monde universitaire, des publications majeures avaient lieu concernant les réseaux de neurones récurrents(Amari, 1972; Schuster & Paliwal, 1997), bien que les premières recherches sur les réseaux de neurones datent de 1800 (Schmidhuber, 2022). Un tournant majeur est survenu au début des années 2010 avec l'avènement des réseaux de neurones profonds et les capacités de calculs permettant leur entraînement. Des modèles tels que Word2Vec (Mikolov et al., 2013), développé par Google en 2013, ont introduit des concepts clés comme l'embedding de mots dans des espaces vectoriels. Word2Vec a considérablement amélioré la capacité des machines à comprendre la sémantique des mots et des phrases. Bien que développé par une entreprise privée, Google a rendu publique la bibliothèque associée à Word2Vec, favorisant ainsi un certain niveau de collaboration et d'accessibilité. La dernière grande étape est l'introduction des architectures de type Transformer (Vaswani et al., 2017), tel que BERT (Bidirectional Encoder Representations from Transformers) (Devlin et al., 2019) par Google en 2018 a révolutionné la compréhension du contexte en NLP. Cette innovation est venue de l'entreprise privée et a jeté les bases d'une nouvelle ère dans le traitement automatique du langage. Puis, les modèles tels que GPT-2 et GPT-3 (Brown et al., 2020) de OpenAI, ont suivi. Dans l'ensemble, l'histoire des modèles de langage a été marquée par une combinaison d'efforts provenant de chercheurs universitaires, d'entreprises privées et de gouvernements. Les contributions open source ont joué un rôle crucial en rendant les connaissances et les outils accessibles à un public plus large, tandis que les entreprises privées ont apporté des innovations clés grâce à des ressources considérables et à des investissements en recherche et développement (R&D).

---

[11] [AltaVista - Wikipedia](#)

[12] [Yahoo! - Wikipedia](#)

## 2. L'IA d'aujourd'hui, une menace pour la gouvernance commune ?

La section précédente a exposé deux aspects cruciaux : d'une part, les communs numériques présentent des caractéristiques uniques qui leur permettent d'éviter la "tragédie des communs"[13], et d'autre part, l'écosystème open source a largement contribué à la rapide évolution de l'intelligence artificielle (IA) mais également à son maintien, car comme nous l'avons détaillé avec ONNX et PyTorch, devenue open-source par la suite. Néanmoins, comme le souligne l'histoire des modèles de langage mentionnée précédemment, les interactions avec les États et les grandes entreprises technologiques sont omniprésentes, et les orientations imposées par ces intérêts privés ou étatiques ne sont pas nécessairement alignées sur les intérêts de la communauté. De plus, l'expansion exponentielle du nombre de paramètres et de la taille des jeux de données requis pour former un modèle de langage a confiné la gestion de ces modèles entre les mains des grands acteurs tels que les GAFAM (Google, Apple, Facebook, Amazon, Microsoft) ou BATX (Baidu, Alibaba, Tencent et Xiaomi,). Cette section examinera deux tendances inquiétantes concernant la gouvernance des modèles de langage : d'une part, la taille croissante de ces modèles a cantonné une grande partie de la communauté dans un rôle de simple consommateur, et d'autre part, nous examinerons les stratégies de certaines nations et conglomérat d'entreprise qui semble se détourner de l'open-source et prendre un virage propriétaire.

### Du rôle d'acteur à celui de consommateur de LM

Le passage d'acteurs à consommateurs de modèles de langage reflète une réalité incontournable dans le paysage de l'IA. Engendrée par l'évolution exponentielle sur le plan de la taille et de la complexité des modèles, cette transition pose des défis croissants pour les nouveaux acteurs désireux de rivaliser avec les ressources massives des GAFAM. Les cycles d'innovation rapides imposés par ces géants technologiques deviennent souvent difficiles à suivre pour le reste de la communauté, laissant ainsi les chercheurs indépendants dans un état de réaction permanente. Pour illustrer cette disparité, prenons l'exemple de la comparaison entre le modèle BLOOM et ChatGPT. BLOOM (Workshop et al., 2023) est un modèle de langue développé par le "BLOOM project" réunissant plus de 1000 chercheurs autour du monde, est une tentative de fournir une alternative open source aux modèles de langage massifs tels que GPT-3. Conçu pour être transparent et accessible, BLOOM est un modèle entraîné avec une vision claire de la reproductibilité, encourageant la collaboration et l'expérimentation dans la communauté. À l'opposé, ChatGPT, développé par OpenAI, a gagné en popularité en raison de sa facilité d'utilisation, mais son opacité suscite des préoccupations en matière de gouvernance et de responsabilité. Le travail qu'a fait OpenAI, sur l'interface utilisateur, l'adaptation du modèle de langue à un besoin bien précis et l'emploi d'évaluateurs humains pour améliorer le modèle ne faisait pas partie des objectifs du BLOOM Project. Aujourd'hui cette initiative n'est que très peu connue du grand public et peu utilisée en comparaison à GPT-3 et chat-GPT.

---

[13] [Tragédie des biens communs — Wikipédia](#)

Enfin, une autre facette expliquant la perte de gouvernance de la communauté envers les modèles de langage réside dans la centralisation des ressources et des données. Alors que les modèles de langage grandissent en taille, le coût d'entraînement et d'infrastructure augmente de manière significative, ce qui limite leur accessibilité à quelques acteurs disposant de ressources considérables. Cette centralisation renforce l'emprise des grandes entreprises et des acteurs puissants sur le développement de ces modèles, compromettant ainsi la diversité et la pluralité des contributions. Récemment, quelques voix se sont élevées pour dénoncer cette centralisation des pouvoirs[14][15], notamment Alexandr Madry, lors de son entretien face au Congrès américain : "*very few players will be able to compete, given the highly specialized skills and enormous capital investments the building of such systems requires.*"[16][17] Ce manque de transparence crée sur le long terme, une crise de la reproductibilité et une fragilité de tous les acteurs économiques qui utilisent comme fondation les modèles de langue fournis par les grandes entreprises.

---

[14] Andrew Ng Warns of Centralized AI Power | by Synced | SyncedReview | Medium

[15] The Paradox of AI Centralization: Are We Building a Titan or a Tyrant? | by BluShark Media | Jul, 2023 | Medium

[16] Advances in AI: Are We Ready For a Tech Revolution? - United States House Committee on Oversight and Accountability

[17] Traduction: "très peu d'acteurs seront en mesure de rivaliser, étant donné les compétences hautement spécialisées et les énormes investissements en capital que la construction de tels systèmes exige"

**Le virage propriétaire de l'IA**

L'intelligence artificielle (IA) suit une trajectoire qui rappelle celle du logiciel libre, se dirigeant de plus en plus vers le modèle mercantile. L'évolution des grands projets open source, tel que le système d'exploitation Linux, offre un exemple éclairant de cette dynamique. Les entreprises commerciales ont rapidement compris l'importance de sécuriser les talents en IA, amorçant depuis 2015 une compétition féroce pour attirer les experts en informatique à la pointe de l'apprentissage automatique et du deep learning (*State Of Venture 2021 Report*,)). Ce phénomène s'apparente aux stratégies d'acquisition de talents adoptées dans le domaine du logiciel libre. Tout comme les firmes ont soutenu financièrement des contributeurs pour les projets ''open-source, les géants de la technologie rivalisent aujourd'hui pour recruter les cerveaux les plus brillants de l'IA. Dans certains cas, les représentants de ces entreprises occupent des postes au sein de fondations open source, participant ainsi aux décisions stratégiques, comme au sein du conseil d'administration de la Fondation Linux. De plus en plus, les entreprises commerciales optent pour l'acquisition de grands fournisseurs de solutions open source prospères, ces derniers générant des revenus considérables grâce à des services tels que l'installation, la formation et le support technique. Un exemple marquant est l'acquisition par IBM de Red Hat, un distributeur majeur de solutions open source dont sa propre version de Linux, pour une somme de 34 milliards de dollars à la fin de l'année 2018. De même, en 2018, Microsoft, porte-étendard des logiciels propriétaires rachète de GitHub, une plateforme collaborative largement utilisée par les développeurs open source pour 7.5 milliards de dollars[18].

Cependant, ce rapprochement entre les domaines du logiciel libre et de l'IA soulève des questions essentielles quant à l'avenir de l'innovation et de la gouvernance au sein de la communauté. Tout comme l'acquisition de startups par les grandes entreprises dans le domaine du logiciel libre a suscité des préoccupations quant à la diversité des approches et à la concentration des pouvoirs, la compétition pour les talents en IA pourrait produire des effets similaires. Les chercheurs et les experts indépendants pourraient se retrouver en marge de cette compétition effrénée, compromettant ainsi la pluralité des idées et des perspectives nécessaires pour une véritable innovation.

---

https://www.latribune.fr/technos-medias/internet/pourquoi-le-rachat-de-github-par-microsoft-pour-7-5-milliards-de-dollars-choque-internet-780682.html

Ce parallèle, nous invite à réfléchir attentivement aux leçons tirées de l'évolution du logiciel libre. Il faut trouver un équilibre entre les intérêts commerciaux et l'ouverture collaborative pour garantir que les ressources liées à l'IA restent une technologie accessible à tous et profitant au plus grand nombre. Nous pensons que les communs peuvent être une réponse à ce défi.

### 3. Ré-inventer l'IA en "commun", l'enjeu de demain.

La montée en puissance du capitalisme de l'IA, où règne une intense concurrence et une concentration des pouvoirs et des moyens a été décrite dans la partie précédente. Ce phénomène d'appropriation d'un bien commun par des entreprises rappelle ce que Bollier appelait "enclosure of the commons". Ce terme, proposé par (Bollier, 2014), désigne une situation dans laquelle les intérêts des entreprises s'approprient notre richesse commune et la transforment en marchandises privées coûteuses. Cela se produit également dans la sphère numérique, où les plateformes contrôlent l'accès aux données et enferment de plus en plus le monde numérique dans leur sphère privée. Pour résister à cette situation, il faut proposer des alternatives, mais également souligner l'importance des données et de l'IA en tant que biens publics, produits par la société et ses membres (Taylor, 2016; Viljoen, 2021). Repenser l'IA comme un commun implique de penser les données comme des communs, également les infrastructures de calcul et enfin les ressources humaines nécessaires à leur développement. Dans ce contexte, l'examen du cas de Hugging Face, un acteur clé dans l'écosystème des modèles de langage, va nous permettre de montrer les limites de l'open-source quant à la gouvernance commune.

**L'initiative "Hugging Face"**

C'est dans ce contexte qu'émerge Hugging Face une start-up franco-américaine, en 2016[19]. Hugging Face se décompose en deux services principaux, une bibliothèque python "Transformers" comportant des implémentations open-source d'architecture de modèle de type Transformers (Wolf et al., 2019). Le second est une plateforme (Hugging Face Hub) d'hébergement de code, de modèles, de jeux de données et d'application web. En 2023, le hub comptait : 120 000 modèles, 20 000 ensembles de données, 50 000 applications de démonstration (Spaces), des centaines d'espaces de discussions (forum) et des dizaines de formations. L'ensemble de ces outils et données sont en libre accès. Chaque utilisateur peut créer un espace de discussion, partager un modèle, créer une communauté dans laquelle des rôles sont assignés et des règles sont décrites. L'entreprise a joué un rôle important dans la publication et la construction de BLOOM, elle a notamment participé à la mise au point de la licence RAIL[20]. Cette nouvelle licence est particulièrement adaptée aux partages de modèle d'IA, car elle limite les usages malveillants. Cette initiative, qu'on pourrait considérer comme le Wikipédia de l'IA, favorise la diffusion des connaissances, des modèles et des données associées à l'IA. Elle encourage les utilisateurs à établir des code de conduites pour chaque communauté[21], elle met à disposition des instruments pour estimer les externalités négatives liées à l'entraînement des modèles[22], avec des mesures précises de l'empreinte carbone et des outils pour détecter les biais de genre ou d'ethnies[23] dans les modèles de langues. Enfin, des solutions "No code" offrant une accessibilité pour tous sont disponibles. Bien que cette initiative soit la plus proche de la définition de commun d'Elinor Ostrom, Hugging Face n'est pas une fondation à but non lucratif et les récents partenariats fait avec Amazon Web Service pour les infrastructures de calcul, ainsi que les investissements conséquents de Google, Amazon, Nvidia, IBM et Qualcomm dans leur dernière levé de fond, nous permettent de douter quant à l'indépendance future de Hugging Face vis-à-vis des géants de la technologie.

---

[19] Hugging Face - Wikipedia

**Penser l'IA en commun.**

Comme le montre l'exemple de Hugging Face, réinventer l'IA en "commun" nécessite une réflexion sur les structures de propriété et de gouvernance. Repenser l'IA en "commun" requiert une conceptualisation des "communs de données" et un investissement dans des capacités de calcul partagées. Comme McDonnell (2018), le définit, les "communs de données" sont des ressources partagées permettant aux citoyens de contribuer, d'accéder et d'utiliser les données comme un bien commun. De nombreux travaux inspirants sur les données en commun, proposent des solutions en matière d'infrastructures de données (Coyle & Diepeveen, 2021). Les "Data trust" sont des exemples d'infrastructure de partage de données dans lesquelles le contrôle des données est transféré à un tiers, qui peut utiliser les données à des fins prédéfinies. Les trusts de données peuvent utiliser des données provenant de différentes sources et permettent de gérer l'utilisation des données pour tous. La solidarité des données est un élément important de la gouvernance, ce qui signifie que les actionnaires des données publiques et des entreprises partagent les avantages et les risques liés à l'accès aux données et à leur production (Delacroix & Lawrence, 2019). La mise en place d'un système de partage et d'accès aux données n'est pas seulement bénéfique pour la société ; elle est également nécessaire à l'innovation dans le domaine de l'IA (*Growing the Artificial Intelligence Industry in the UK*, 2017.). Parallèlement, investir dans des capacités de calcul communes peut contribuer à démocratiser l'IA, en impliquant davantage d'individus et d'organisations dans son développement. Cela pourrait nous aider à devenir moins dépendant de l'infrastructure privée des GAFAM. Alors que le secteur des entreprises prétend souvent que l'investissement public étouffe l'innovation; (Mazzucato, 2015) réfute ce mythe et affirme en fait que les technologies à l'origine, par exemple, de l'iPhone (GPS, écran tactile et Siri) ont toutes été soutenues par des fonds publics. Un autre exemple est l'algorithme de recherche de Google, qui a été financé par des fonds publics par l'intermédiaire de la "National Science Foundation" (NSF). Ainsi, nous pensons que les gouvernements devraient mettre davantage en commun les capacités de calcul. De la même manière qu'il existe une politique pour l'open data en France avec la plateforme Data gouv , nous pourrions imaginer une politique de mise à disposition des ressources de calculs.

---

[20] Responsible AI Licenses (RAIL)
[21] Code of Conduct – Hugging Face
[22] CO2 Emissions and the 🤗 Hub: Leading the Charge
[23] DataMeasurementsTool - a Hugging Face Space by huggingface

L'exemple d'HuggingFace, offre des pistes prometteuses pour réinventer l'IA dans un esprit de partage, de collaboration et d'intérêt collectif, mais il est nécessaire que les services publics s'emparent de ce sujet afin de fournir les moyens nécessaires à faire de l'IA un atout pour l'ensemble des citoyens.

**Conclusion**

En retraçant l'histoire des modèles de langues et en s'appuyant sur la définition de commun tel que proposé par Elinor Ostrom, nous avons tenté de décrire la centralisation des pouvoirs dans l'écosystème IA. Cette centralisation semble être une menace pour la gouvernance commune alors que ces ressources numériques deviennent des outils de notre quotidien. Bien que des initiatives prônant la culture du logiciel libre existent (Hugging Face), de nombreux efforts restent à faire concernant la gouvernance des données, des infrastructures de calcul et des moyens humains afin que l'IA soit bénéfique à l'ensemble de la société. Les conséquences de l'IA sur nos vies, mais également sur le changement climatique, impliquent que les prises de décisions concernant les directions futures soient le fruit d'une réflexion collective et non plus seulement guidée par l'intérêt financier des grands groupes technologiques.

# Références